\documentclass[journal=jacsat,manuscript=article]{achemso}

\usepackage[version=3]{mhchem} 
\usepackage{paralist}
\usepackage{booktabs,siunitx}
\usepackage{array}
\usepackage{dcolumn}

\newcolumntype{H}{>{\setbox0=\hbox\bgroup}c<{\egroup}@{}}
\usepackage{amssymb}

\usepackage[flushleft]{threeparttable}
\usepackage[T1]{fontenc}
\usepackage{booktabs,braket,cleveref,mathptmx,xcolor,siunitx,dcolumn,graphicx,bm,graphics,caption,subcaption,multirow,array,paralist,rotating}
\usepackage[normalem]{ulem}%

\newcommand{\abinitio}{\emph{ab initio}}
\newcommand{\cm}{cm$^{-1}$}
\newcommand{\um}{$\mu$m}

\newcolumntype{d}{D{.}{.}{-1}}
\newcommand{\totmol}{2914}
\newcommand{\calcmol}{2743}
\newcommand{\harmdsname}{CHNOPS\calcmol-HARMONIC}
\usepackage{rotating}

\graphicspath{{Figs/}}

\SectionNumbersOn

\author{Juan C. Zapata Trujillo}
\author{Maria M. Pettyjohn}
\author{Laura K. McKemmish}
\email{l.mckemmish@unsw.edu.au}
\affiliation[University of New South Wales]
{School of Chemistry, University of New South Wales, 2052, Sydney}

\title[An \textsf{achemso} demo]
  {High-throughput Quantum Chemistry: Empowering the Search for Molecular Candidates behind Unknown Spectral Signatures in Exoplanetary Atmospheres}

\usepackage[utf8]{inputenc}
\DeclareUnicodeCharacter{2212}{-}
\begin{document}







\begin{abstract}
    The identification of molecules in exoplanetary atmospheres is only possible thanks to the availability of high-resolution molecular spectroscopic data. However, due to its intensive and time-consuming generation process, at present, only on order 100  molecules have high-resolution spectroscopic data available, limiting new molecular detections.
    
    Using routine quantum chemistry calculations (i.e., scaled harmonic frequency calculations using the B97-1/def2-TZVPD model chemistry with median errors of 10\,\cm{}), here we present a complementary high-throughput approach to rapidly generate approximate vibrational spectral data for 2743 molecules made from the biologically most important elements C, H, N, O, P and S.  Though these data are not accurate enough to enable definitive molecular detections and does not seek to replace the need for high-resolution data, it has powerful applications in identifying potential molecular candidates responsible for unknown spectral features. We explore this application for the 4.1\,\um{} (2439\,\cm{}) feature in the atmospheric spectrum of WASP-39b, listing potential alternative molecular species responsible for this spectral line, together with \ce{SO2}. Further applications of this big data compilation also include identifying molecules with strong absorption features that are likely detectable at quite low abundances, and training set for machine learning predictions of vibrational frequencies.
    
    Characterising exoplanetary atmospheres through molecular spectroscopy is essential to understand the planet's physico-chemical processes and likelihood of hosting life. Our rapidly generated quantum chemistry big data set will play a crucial role in supporting this understanding by giving directions into possible initial identifications of the more unusual molecules to emerge.
\end{abstract}


\section{Introduction}

Emerging data from the James Webb Space Telescope (JWST) is providing extraordinary opportunities to deepen our understanding of the molecular catalogue of exoplanetary atmospheres. For example, recent single-transit observations of the infrared spectrum of WASP-39b (a hot-gas-giant exoplanet) have led to the first clear atmospheric exoplanetary detection of \ce{CO2} and photochemically produced \ce{SO2} \citep{23AlWaAl}. 

The detection of molecular species in exoplanetary atmospheres is enabled by the availability of high-resolution and high-completeness molecular spectroscopic data \citep{ExoMol2020, 21Mc,22GoRoHa,20Be}, usually line lists, providing a comprehensive set of energy levels and transition intensities. It is through the synergy between highly accurate experimental measurements and high-level \abinitio{} quantum chemistry calculations that that we can consolidate this high-resolution spectroscopic data for a given molecule.  Standard line lists repositories storing high-resolution and high-completeness spectroscopic data for different molecules include HITRAN/HITEMP \citep{22GoRoHa}, ExoMol \citep{12TeYu,exomol}, and MoLLIST \citep{20Be} which have all been crucial in providing the necessary data for enabling molecular observations in different astronomical bodies \citep[e.g.,][]{23GrDiTa,23KoAbDi,23NiRaCo,22SiEdZi,21NuKaGi,21BrTaCh}. 

Producing and consolidating high-resolution spectroscopic data is, however, an extensive and time-consuming process, as evidenced by the low number of molecules with current line lists available (approximately  100), thus limiting the scope of new molecular observations. A perfect example showcasing this limitation is the unassigned 4.25\,\um{}  feature in the transmission spectrum for WASP-39b, where none of the available molecular line lists could reproduce the correct wavelength and bandwidth for this unknown spectral signal \citep{23AlWaAl}. This limitation will become more acute in coming years as more detailed observations are made.

Understanding the complexity behind generating high-resolution spectroscopic data and its current molecular coverage limitations, \cite{19SoPeSe} pioneered a complementary big data approach (called Rapid Approximate Spectral Calculations for ALL or RASCALL) for generating approximate vibrational spectroscopic data for a large number of molecules rapidly.  RASCALL is a computational approach relying on the fact that a molecule's infrared (vibrational) spectrum is mainly influenced by the localised vibrations of the functional groups present in the molecule (i.e., the specific groups of atoms and bonds responsible for the molecule's physico-chemical properties, e.g. alcohols (\ce{-OH}), carbonyl (\ce{C=O}), etc.). This approach generates, within seconds, a good first approximation to each molecule's infrared spectra, including qualitative transition intensities for the corresponding band centres.

RASCALL is a powerful tool with appealing anticipated applications in astrochemistry \citep{19SoPeSe}. Most notably, its underlying data can be crucial in examining regions with high spectral congestion, identifying areas of uncertainty in molecular detections and guiding future astronomical observations towards less-convoluted spectral regions. Similarly, comparisons between different spectra produced by RASCALL can provide informed directions into protocols prioritising high-resolution experimental measurements and high-level \abinitio{} quantum chemistry calculations for different molecular species. This analysis can help alleviate the selection process for bridging the gap of molecules with no line lists available.

However, RASCALL's functional-group-theory foundation has some inherent limitations that are worth addressing \citep{19SoPeSe,21ZaSyRo}. Most notably, RASCALL does not take into account the influence that the chemical environment around functional groups has in the predicted frequencies, nor is capable of calculating frequencies in the fingerprint spectral region (below 1000\,\cm{}), mostly corresponding to global (involving large parts of the molecule) instead of functional group (localised) vibrations. For molecules with similar structures, these limitations imply significant similarity in their RASCALL spectra, leading to potential misassignments and molecular ambiguities, thus restricting the broad applicability of RASCALL.

Recognising the potential of this bulk approximate spectral data to support exoplanet atmosphere molecule identification, here we present an alternative methodology utilising standard computational quantum chemistry approaches (instead of high-level \abinitio{} quantum chemistry) that addresses the key limitations of RASCALL. Specifically, we pioneer a high-throughput largely automated approach to calculate approximate vibrational spectral data for thousands of molecules of astrochemistry interest with an expected median error on our approximate spectral data of 10\,\cm{} ($\sim$\,0.02\,\um{} around 2500\,\cm{}). These calculations and their underlying generated data are backed up by a thorough benchmark study \citep{23ZaMc}. Unlike with RASCALL, our approach naturally allows prediction of the influence of chemical environment on functional group spectral positions, frequencies in the fingerprint region, and quantitative intensity predictions. Extensions to predicting overtone and combination band frequencies, rotational profiles of vibrational bands, and high-energy conformers are also feasible within this framework.  

Though our quantum chemistry data are not accurate enough to enable definitive molecular detections in exoplanetary atmospheres, the data nevertheless allows identification of potential molecular candidates for unknown spectral signals. We showcase and discuss this powerful application in Section \ref{sec:astrochem} of this manuscript using the \ce{SO2} detection on WASP-39b \citep{23AlWaAl} as an example.

Our primary goals for this paper are to (1) demonstrate how approximate spectral data can help molecular identification in exoplanet atmospheres and (2) develop a largely automated, high-throughput approach to produce these data, highlighting its versatility.  We provide as supporting information the new approximate spectral data for predicted fundamental frequencies and transition intensities of \calcmol{} molecules (referred to as the \harmdsname{} dataset), for use in both astrochemistry and other applications. These data can also be accessed through the Harvard Dataverse repository at \url{https://doi.org/10.7910/DVN/0DLSDP}.

To achieve these goals, this paper is organised as follows:

\begin{description}
    \item Section \ref{sec:chnops} outlines the list of potential biosignatures (the \harmdsname{} dataset) for which we produce new data, describing their elemental and functional group distributions. 
    \item Section \ref{sec:QC} provides a thorough introduction for non-specialists to the key computational quantum chemistry concepts describing our approach and delineates the expected errors in our produced approximate spectral data. We also provide references to useful reviews and textbooks for the interested reader. 
    \item In Section \ref{sec:approach} we present our novel high-throughput quantum chemistry methodology to produce approximate vibrational spectral data for all molecules in our working set, detailing the different steps in the calculation's pipeline. 
    \item Section \ref{sec:vib_spec} presents and discusses the generated vibrational data, making individual comparisons with experiment. 
    \item In Section \ref{sec:astrochem} we delve into the applications of our quantum chemistry big data in the context of the recent \ce{SO2} detection on WASP-39b, highlighting potential alternative molecular candidates. 
    \item Finally, in Section \ref{sec:final_remarks}, we outline the promise of our approach, its limitations, and future directions of research. 
\end{description} 

\section{The \harmdsname{} dataset of potential biosignatures}
\label{sec:chnops}

Noting the variety of molecules produced by life on Earth and the large diversity of exoplanetary environments, \citet{16SeBaPe} collated a list with over 16,000 molecules that could be initially considered as potential biosignature gases if detected in exoplanetary atmospheres. These are volatile molecules with up to six non-hydrogen atoms that are stable under standard temperature and pressure conditions (STP). Note that different isomers are considered distinct molecules within the algorithm used by \citet{16SeBaPe}, with unique SMILES codes. 

We start our compilation of approximate vibrational spectra with a subset of this list, selecting  molecules containing C, H, N, O, P, and S only as these are the most important elements biologically. The full subset of potential biosignatures contains \totmol{} molecules; however, our high-throughput automated approach (see Section \ref{sec:approach}) only produced approximate vibrational spectral data for \calcmol{} molecules with up to 20 total atoms and 6 total non-hydrogen atoms.

\begin{figure}
    \centering
    \includegraphics[width = 0.8\textwidth]{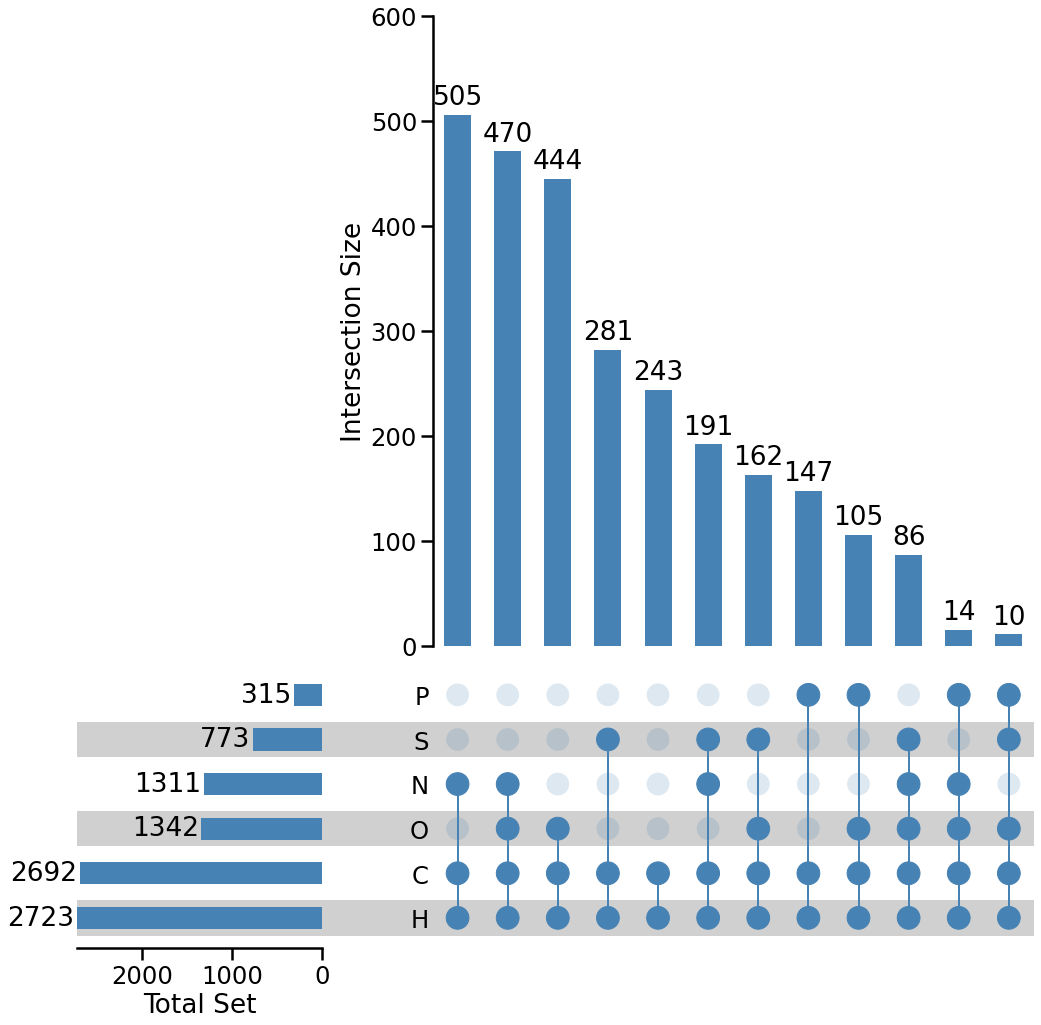}
    \caption{Distribution of elements within \harmdsname{}. The horizontal histogram on the left details the number of molecules that contain the corresponding element to the right. The dots create sets of elements that form various molecules, with the counts of those sets shown in the histogram above, e.g., the first column in the dots diagram contains 505 molecules that are made up of N, C, and H only. Subsets of elements with intersection size of fewer than 10 molecules have been excluded for simplicity.}  
    \label{fig:atoms}
\end{figure}

\Cref{fig:atoms} presents the counts for (1) the element composition (left-side rotated histogram) and (2) multiple combinations of elements (top histogram) for all molecules in the \harmdsname{} dataset. The central dots-diagram in the figure presents the sets of elements that form various molecules as illustrated by the connected filled-dots, e.g., the first three connected filled-dots in the left-side of the diagram represent molecules containing C, H, and N only, and the histogram bar above shows that this particular combination of elements adds up to 505 molecules in our working set. The left-side rotated histogram shows that C and H are the most abundant elements in the list, but there are actually more molecules containing C, H, and N (505 in total), than C and H only (243). P-containing molecules represent the least common molecular type in our working set.

To further understand the chemical features underpinning our working set of molecules, we used a python code based on the RDKit library\footnote{RDKit: Open-source cheminformatics; \url{http://www.rdkit.org}} to obtain approximate counts for the most common functional groups present in all molecules in the \harmdsname{} dataset. We categorised all functional groups into five main classes following the elemental composition presented in \Cref{fig:atoms}: hydrocarbons (molecules with C and H only), N-containing groups, O-containing groups, S-containing groups, and P-containing groups.

\begin{figure}
    \centering
    \includegraphics[width=0.8\textwidth]{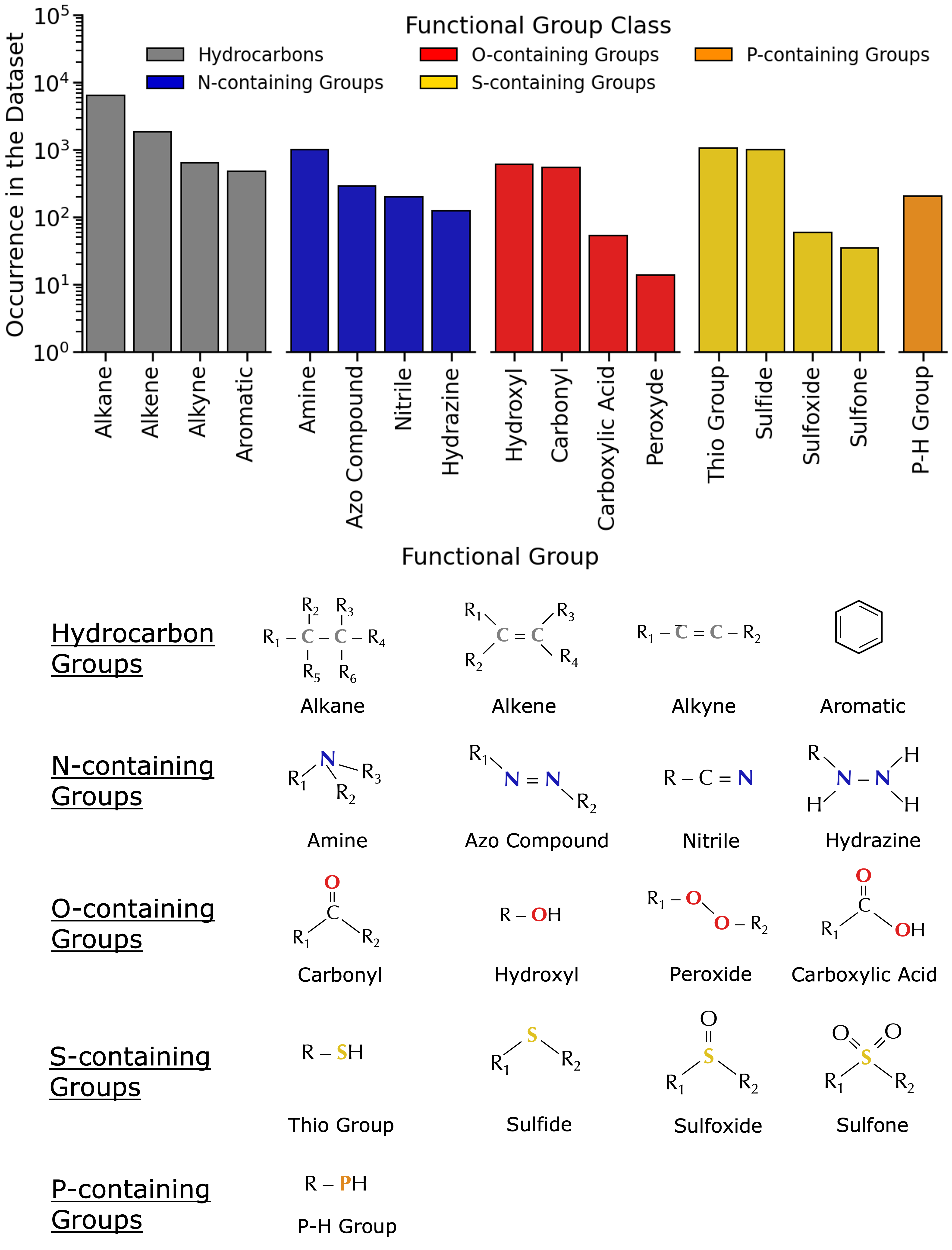}
    \caption{Logarithmic-scale distribution of the most common functional groups in the \harmdsname{} sorted on their elemental composition. A schematic representation of the functional groups is presented at the bottom of the histogram. $R_{n}$ is used to denote C-containing groups or chains.}
    \label{fig:fgroups}
\end{figure}

\Cref{fig:fgroups} showcases the occurrence (in logarithmic scale) of the most common functional groups within each class in the \harmdsname{} dataset. In line with the data reported in \Cref{fig:atoms}, hydrocarbon functional groups are the most widespread across our working set of molecules. Amines, thiols, and alcohols also have significant abundance. 

\section{Computational Quantum Chemistry} 
\label{sec:QC}

Our goal in this paper is to rapidly and automatically generate approximate vibrational spectral data for thousands of molecules of astrochemistry interest, specifically the set of CHNOPS potential biosignatures described above. The very high-accuracy quantum chemistry approaches traditionally used in astrochemistry, e.g., molecular line list creation, are too computationally expensive to be feasible and so we rely on standard and less computationally demanding approaches for our quantum chemistry big data generation. 

Fortunately, using largely standard approaches, modern computational quantum chemistry (specifically scaling harmonic frequencies using the B97-1 method \citep{98HaCoTo} and def2-TZVPD basis set \citep{05WeAh,10RaFu}) enables routine prediction of fundamental vibrational frequencies (i.e., infrared spectra) of small to medium-sized molecules to a median accuracy of approximately 10\,\cm{}, equivalent to 0.01--0.1\,\um{} depending on the wavelength considered \citep{23ZaMc}. The median error is largely independent of molecule size and frequency range, though of course calculation time increases significantly for larger molecules. 

We start with a summary of the important concepts of modern computational quantum chemistry approaches in the astrochemistry context, going through methods to predict molecular vibrational spectra with the double harmonic approximation or anharmonic approaches based on the lowest-energy molecular geometry only. We then provide a description of the method or level of theory (approximation to the electronic Schr\"odinger equation) and the basis set (mathematical functions used to describe electron distribution), which collectively known as the model chemistry, and control the accuracy of the calculation. 
We fully justify our choice of model chemistry (B971/def2-TZVPD) in the final sub-section here, including a thorough discussion of its anticipated error profile. 

\subsection{Predicting Vibrational Frequencies}

The bands observed in a molecule's infrared spectrum correspond to excitations between different vibrational energy levels within the molecule. Therefore, to accurately predict the molecular infrared spectra, we must model the vibrational energy levels in the molecule, as well as the strength of the transitions between these levels (vibrational transition intensities).

Line list constructors most commonly find these vibrational energy levels and transition strengths through direct solution of the nuclear motion Schr\"odinger equation based on a full potential energy surface; for examples, see \cite{Duo,TROVE}. This approach yields a very large number of quantised vibrational energy levels and wavefunctions, with transition intensities computable through the use of the dipole moment surface and explicit wavefunctions. This full solution, however, is extremely time-consuming and limited to very small molecules like \ce{CH4}. 

Within the broader chemistry community, a far less time-consuming approach is employed to simulate infrared spectra of very large molecules based solely on derivatives calculated at the minimum-energy geometry of the molecule. It is these approaches that are practical to consider for our high-throughput methodology.

\subsubsection{Physical Chemist's Perspective of Vibrational Spectroscopy}
Transitions between vibrational energy levels in a molecule can be most simply described through different populations of normal modes, which are collective motions of all the atoms in a molecule. 

The fundamental transitions are the strongest lines in infrared spectra and arise from excitations from the vibrational ground state (with zero quanta in all vibrational normal modes) to vibrational states with just one quantum of excitation in one vibrational mode. Predicting the frequency and intensity of these fundamental transitions is an important task in quantum chemistry and is sufficient for most predictions of infrared spectra in chemistry. 

In vibrational spectroscopy, patterns emerge that allow chemists to predict infrared spectra of molecules above 1000\,\cm{} without calculations. High-frequency normal modes often involve localised vibrations, such as bond stretches or angle bends. The frequencies of these local modes primarily depend on the component atoms with minor variations based on the chemical environment. This principle informs RASCALL's functional-group approach. 

In contrast, the low-frequency modes (below 1000\,\cm{}) involving collective motions of large parts of the molecule cannot be easily predicted by chemists or RASCALL, with computational quantum chemistry methods needed for even qualitative accuracy. This region is called the ``fingerprint" region and is valuable for definitive identification of unknown molecules.

Transitions involving two quanta of excitation in a single mode (overtone band) or two modes (combination band) are typically weaker but can be significant in astronomical observations, e.g. the 4.1 \um{} \ce{SO2} band in WASP-39b (see Section \ref{sec:astrochem}). Anharmonic calculations are needed to predict these bands, available in certain quantum chemistry packages.

Excitations from already excited vibrational states are known as hot bands and become more intense at higher temperatures. Predicting their frequencies is possible with anharmonic calculations, but determining intensities requires additional processing beyond typical quantum chemistry packages.

Prediction of vibrational band centres, e.g., frequencies for fundamental, overtones or combination bands, through computational quantum chemistry does not consider any rotational effects. However, when rotational motion is considered, the band becomes significantly broadened with the band peak frequency often shifting away from the band centre. This broadening occurs because of the significant population of excited rotational states, and simultaneous excitation or de-excitation of rotational quantum levels alongside the vibrational excitation.

Our \harmdsname{} dataset contains only predictions of fundamental frequencies. Our analysis in \Cref{sec:astrochem} necessitates limited calculation of overtone and combination band frequencies. We do not consider hot bands in this paper.

\subsubsection{The Double Harmonic Approximation}

The standard approach to calculating infrared (IR) spectra is the double harmonic approximation in which both potential energy and dipole moment surfaces are expanded as a Taylor series truncated to the second order (i.e., a harmonic potential) with the force constant (Hessian) matrix calculated at the minimum of the potential energy surface \citep{06Tu}.

Standard quantum chemistry packages can readily predict vibrational frequencies and transition intensities for the fundamental modes within the double harmonic approximation. 

However, harmonic vibrational frequencies tend to overestimate experimental fundamental frequencies \citep{81PoScKr}. Fortunately, this overestimation is largely systematic and applying scaling factors to calculated harmonic frequencies results in improved vibrational fundamental frequency predictions. A comprehensive review on the use of scaling factors for predicting fundamental vibrational frequencies through scaling harmonic frequency calculations can be found in \cite{21ZaMc}, with up-to-date recommendations outlined in \cite{23ZaMc}.

\subsubsection{Anharmonic Approaches}

One of the main limitations of the double harmonic approximation is that it only allows the calculation of predicted fundamental frequencies and intensities; in fact, overtones and combination bands are forbidden within the double harmonic approximation. Instead, anharmonic approaches are needed to predict overtones and combination bands, in which the potential energy surface is no longer represented by harmonic potentials but takes on a more complex form allowing a better physical representation of the potential energy surface for the molecule considered. Direct prediction of fundamental vibrational transitions are also possible within anharmonic approaches, though scaling factors are still often used to improve performance \citep{jacobsen2013anharmonic}. Hybrid methods are also possible, where an expensive high-accuracy model chemistry is used within the double harmonic approximation with anharmonic corrections calculated using cheaper model chemistries \citep{biczysko2010harmonic,biczysko2018computational}.

Various anharmonic treatments are available \citep{sibert2023modeling} but vibrational second-order perturbation theory (VPT2) has become the most popular \citep{biczysko2018computational,barone2021computational}, with strong performance for reasonably modest calculation times \citep{14BaBiBl,15BaBiPu,15Bl,18BiBlPu}. Nevertheless, even VPT2 calculations scale much more steeply with molecular size than harmonic approaches as fourth-order derivatives of molecular energy with respect to nuclear coordinates are required, making these calculations practical only for small to medium-sized molecules.

\subsection{The Method or Level of Theory}

In the context of this work, the central principle behind computational quantum chemistry is finding approximate solutions to the non-relativistic time-independent electronic Schr\"odinger equation, $\hat{H}_\textrm{elec}\Psi_\textrm{elec} = E_\textrm{elec}\Psi_\textrm{elec}$, that assumes that the electronic and nuclear motions can be separated due the disparate mass of the electrons and the nuclei (Born-Oppenheimer approximation \citep{27BoOp,10PuStGa}). In this equation, $\hat{H}_\textrm{elec}$ corresponds to the electronic Hamiltonian operator describing the motion of electrons in the fixed nuclear field, $\Psi_\textrm{elec}$ is the electronic wave function, and $E_\textrm{elec}$ is the electronic energy.

Because exact solutions to the electronic Schr\"odinger equation are only possible for one-electron systems (e.g., H$^{+}_{2}$), approximate solutions to many-electron systems, which are indeed the main focus in different chemical applications, need to be found. We call these approximations the method or level of theory and they can be classified into two main families: (1) wavefunction methods and (2) density functional theory (DFT) methods. 

\subsubsection{Wavefunction Methods}

As specified in its name, wavefunction methods focus on finding high-quality  representations of the electronic wavefunction $\Psi_\textrm{elec}$. 

Central to this problem, and setting the foundations to computational quantum chemistry, is the Hartree-Fock (HF) method that places electrons in molecular orbitals and uses a self-consistent field (SCF) procedure to optimise the shape of these orbitals in order to variationally minimise the energy of the electronic wavefunction \citep{szabo2012modern}. 

The quality of the results from the HF method are typically very poor due to the lack of electron correlation, and thus in practice HF is never used (instead for fast calculation, DFT methods are preferred). 

Post-HF methods build on the HF wavefunction by extending the wavefunction ansatz to give additional flexibility to the electronic wavefunction, thereby enabling electron correlation to be well described. For instance, the CCSD(T) method, referred as the gold-standard methodology in quantum chemistry, uses a coupled-cluster ansatz for the wavefunction (the "CC") and incorporates singles and double electron excitations from occupied to unoccupied molecular orbitals variationally (the "S" and "D"), as well as triple excitations perturbatively (the "(T)") \citep{bartlett2007coupled}.

In the context of astrochemistry,  CCSD(T) is often used to generate potential energy and dipole moment surfaces as part of the first steps in the line list construction process. However, CCSD(T) can have significantly reduced accuracy when describing regions of the potential energy surface far from the minimum energy geometry, when modelling transition metal diatomic systems with complex electronic structures (common in hot astronomical bodies like cool stars and hot Jupiters), and cannot be used to describe many excited electronic states. In these cases, multi-reference configuration interaction methods with variationally-considered single and double excitation (typically abbreviated to MRCI) are used \cite{tennyson2016ab}.

Both CCSD(T) and MRCI are very demanding methods with very large memory requirements. Therefore these methods are restricted for the smallest molecular systems, e.g. \ce{CH4} and \ce{TiO}. 

For larger systems, the most accurate modern methods are density functional theory (DFT) methods (noting that cheaper wavefunction methods like HF are often a component of DFT calculations).

Wavefunction methods have generally been well-established for decades, and hence the excellent \cite{szabo2012modern} textbook remains a great reference for their mathematical foundations,  including coupled-cluster approaches. For the very high accuracy multi-reference methods, the recent \cite{MR_review} provides a thorough discussion of MRCI and common alternatives, including a discussion of emerging approaches with significant promise for faster calculations with accuracies similar to those obtained with traditional MRCI methods.

\subsubsection{Density Functional Theory (DFT) Methods}
 
Contrary to wavefunction methods, DFT methods focus on the electron density \(\rho\) rather than the wavefunction to find approximate solutions to the electronic Schr\"odinger equation \citep{65KoSh}. In theory, the energy of any quantum system can be written as a functional of its electron density, but the mathematical form of this function is unknown. A wealth of different approximations to this unknown functional have thus been developed based on different principles (e.g., physically motivated, fitting to experiment, incorporating some wavefunction component to the energy). The choice of DFT methods for particular molecular systems and properties is a crucial area of modern quantum chemistry research.  

In practical terms, DFT methods have higher accuracy than wavefunction methods for their computational time, and have replaced all but the most high-cost, high-accuracy wavefunction methods like CCSD(T) and MRCI. 

Mathematically, within the DFT approach, the energy of the system is a function of the electron density,
\begin{equation}
    E = T[\rho(\textbf{r})] + E_\textnormal{ne}[\rho(\textbf{r})] + E_\textnormal{ee}[\rho(\textbf{r})] +E_\textnormal{xc}[\rho(\textbf{r})],
\end{equation}
where \(T\) is the kinetic energy of the electrons, \(E_\textnormal{ne}\) is the attraction between the nucleus and electrons, \(E_\textnormal{ee}\) is the repulsion between electrons, and \(E_\textnormal{xc}\) is the exchange and correlation energies \citep{65KoSh}. It is the last term, typically called the exchange-correlation functional, that is unknown and the difference between most modern density functional approaches. 

The simplest and cheapest DFT methods are functions only of the electron density and its gradients, but more accurate approaches incorporate in their energies contributions calculated with cheaper wavefunction methods. The two classes of most relevance here are the hybrid functionals, which incorporate HF energies and have computational cost scaling approximately as O($n^3$) ($n$ being the system size such as the number of atoms or basis functions \citep{19GoMe}), and double-hybrid functionals, which use energies computed with second-order perturbation theory (MP2) and have computational scaling of approximately O($n^5$). 

A great introduction to DFT methods for non-experts is provided by \cite{goerigk2019trip}. 

\subsection{The Basis Set}

Both wavefunction methods and most DFT approaches form molecular orbitals ($\phi$) by linear combination of atom-centred basis functions ($\chi$), i.e. $\phi_i = \sum_{j} c_{ij} \chi_j$. The specific basis functions available in a calculation are defined by the user,  with the coefficients $c_{ij}$ optimised within a calculation. 

Molecular basis sets are sets of basis functions defined for many elements across the periodic table. Basis sets are designed to compactly describe electron distributions in molecules and their variation between chemical species. Larger basis sets provide higher accuracy for greater computational costs. 

Vibrational frequencies are an  example of a property that depends primarily on the description of the valence electron region of the system and thus benchmarking shows that standard basis set design approaches have strong performance. Modern basis set design achieves high accuracy for minimal extra cost by increasing the number of basis functions used to describe valence electrons (the `zeta' level of the basis set) and by adding higher angular momentum `polarisation' functions to describe changes in electron distribution due to bonding. Benchmarking by \cite{23ZaMc} shows that for vibrational frequencies, the basis set incompleteness error is roughly the similar magnitude to the density functional approximation error at a double-zeta level with one set of polarisation functions, and smaller for triple-zeta basis sets with two sets of polarisation. 

Vibrational intensities, however, depend on a quality description of the variation in the dipole moment of the molecule across the vibrational motion. Dipole moment is a property that depends on accurate description of the electron density far from the nucleus and benchmarking by \cite{20ZaMc} shows that dipole moment prediction accuracy is significantly enhanced when diffuse functions (with a larger spatial extent) are added to the basis set; this is typically signified by the basis set prefix "aug" or the "D" suffix.

\subsection{Model Chemistry Choice and Anticipated Errors}

\begin{figure}
    \centering
    \includegraphics[width = 0.8\textwidth]{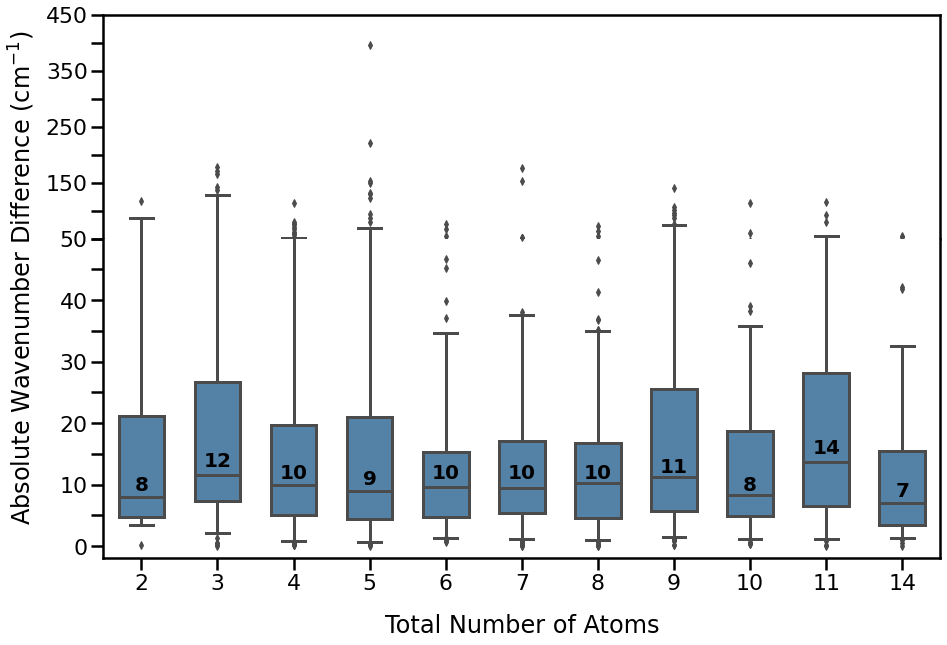}
    \caption{Distribution of errors for the B97-1/def2-TZVPD model chemistry between the predicted fundamental and experimental fundamental frequencies for all 141 molecules in the VIBFREQ1295 dataset as a function of the molecule size. The numbers within the boxes represent the median error for each group considered. The bottom and top whiskers in the figure encapsulate 5 and 95\,\% of the data, respectively. Note that we changed the scale of the absolute wavenumber difference axis for data points above 50\,\cm{} to allow readability of both the boxes and outliers.}  
    \label{fig:error_molsize}
\end{figure}

We generate approximate vibrational spectra within the double harmonic approximation. The model chemistry chosen is the B97-1 hybrid functional method together with the def2-TZVPD basis, a triple-zeta basis set augmented with diffuse functions. This choice is based on extensive benchmarking of computed versus experimental fundamental vibrational frequencies \citep{23ZaMc} across 141 molecules and 1295 frequencies \citep{22ZaMc_VIBFREQ}, and the assessment of predictions of dipole moment by various model chemistries \citep{20ZaMc}. 

Our rationale for this choice is as follows:
\begin{enumerate}
    \item Double-hybrid functionals greatly increased computational cost but didn't significantly improve accuracy;
    \item B97-1 is the best-performing hybrid functional in the benchmark study; 
    \item The inclusion of the diffuse functions in the def2-TZVPD basis set (i.e., the "D") was crucial to increase the accuracy of predicted vibrational intensities; and
    \item Though a double-zeta basis set is likely a sufficient choice for our study, the modest impact on computational time was deemed reasonable to slightly increase calculation accuracy. 
\end{enumerate}

\Cref{tab:mc_calcs} presents statistical metrics for the performance of B97-1/def2-TZVPD model chemistry in harmonic frequency calculations, using a set of 141 molecules similar to our \harmdsname{} set. From these results, we anticipate median errors in our approximate vibrational spectra calculations of 10\,\cm{}, with 75\,\% of calculations having errors less than 19\,\cm{}. Outliers do occur, with about 5\,\% of frequencies having very large errors exceeding 60\,\cm{}; these are unpredictable in general, though open shell systems, like radicals, often have large errors.

\begin{table}
    \centering
    \caption{Statistical metrics for the B97-1/def2-TZVPD model chemistry in harmonic frequency calculations as described in \cite{23ZaMc}. Note that the errors in \um{} are given at different wavelengths.}
    \begin{tabular}{lcccc}
    \toprule
        \multirow{2}{*}{Statistical Metric} & \multicolumn{4}{c}{Value}                                \\
                                            & in \cm{} & At 10\,\um{} & At 5\,\um{}  &  At 3.3\,\um{}  \\
        \midrule
        Median error (50\,\%)               & 10       & 0.10         & 0.03         & 0.01            \\
        75\,\% percentile                   & 19       & 0.19         & 0.05         & 0.02            \\
        95\,\% percentile                   & 60       & 0.60         & 0.15         & 0.07            \\
        \vspace{-0.7em}                  \\
        Mean Absolute Error                 & 17       & 0.17         & 0.04         & 0.02            \\
    \bottomrule
    \end{tabular}
    \label{tab:mc_calcs}
\end{table}

Importantly, the accuracy of our method does not depend on molecule size, as shown in \Cref{fig:error_molsize} for the benchmark data in \cite{22ZaMc_VIBFREQ} (similar types of molecules to the \harmdsname{} set).

\section{Automated High-throughput Spectral Predictions}
\label{sec:approach}

We followed a newly developed automated high-throughput approach to calculate approximate vibrational spectral data for \totmol{} molecules in the full set of CHNOPS potential biosignatures, with successful calculations for \calcmol{} molecules (\harmdsname{}). 

\subsection{High-Throughput Methodology}

\begin{figure}
    \centering
    \includegraphics[width=0.6\textwidth]{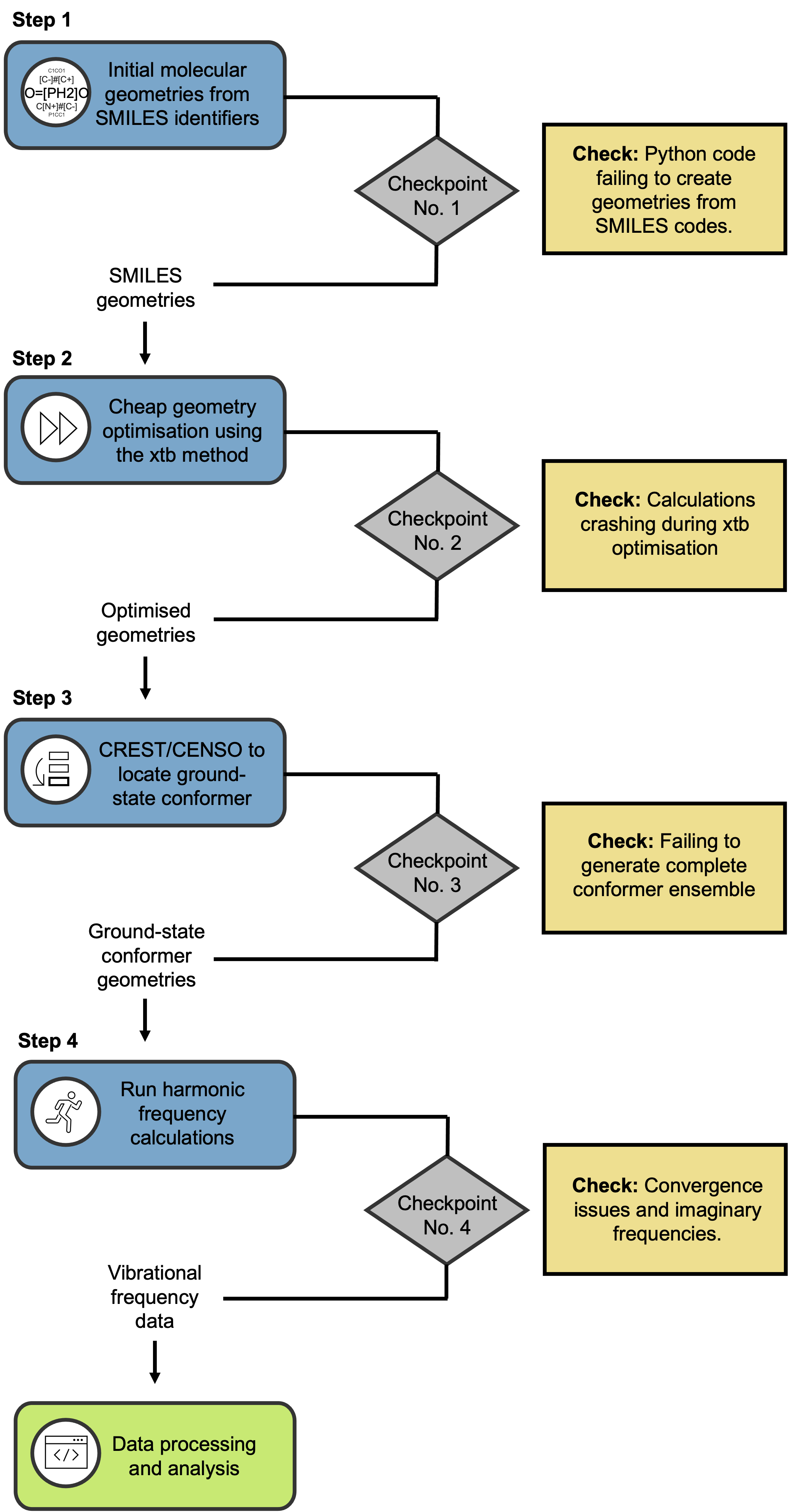}
    \caption{Semi-automated high-throughput approach outlined in this publication. The boxes in blue represent the different steps taken in the data generation process before the corresponding data processing and analysis, which is represented in the green box at the end of the workflow. We had different checkpoints throughout (rhombus in grey) with their description highlighted in the right-side yellow boxes. Solid lines in the workflow represent automated steps where python and bash scripts were used, and dashed lines represent steps requiring further manual handling.}
    \label{fig:HT_Approach}
\end{figure}

\Cref{fig:HT_Approach} presents a flow chart of our high-throughput approach highlighting the different steps (boxes in blue) and checkpoints (rhombi in grey) considered. The yellow boxes describe the actions undertaken at the different checkpoints. 
\newline

\noindent\textbf{Step 1: Initial Molecular Geometries} \newline
We used the RDKit and Chemcoord \citep{weser2023automated} libraries in python to generate initial molecular geometries for all molecules using their Simplified Molecular Input Line Entry System (SMILES) \citep{88We,89WeWeWe,90We}; a line notation describing the structure of chemical species. The algorithm takes the SMILES identifier as an input and performs a crude geometry optimisation using molecular mechanics (i.e., Merck molecular force field (MMFF) \citet{toscoBringingMMFFForce2014}) to output an initial geometry. At the end of this step, we checked for molecules where the python code failed to generate the initial geometries (\textit{checkpoint No. 1}), excluding these molecules from our approach.

\vspace{1em}
\noindent\textbf{Step 2: Cheap Geometry Optimisation} \newline
Using the initial geometries obtained from the SMILES identifiers in the previous step, we performed a cheap geometry optimisation using the GFN2-xTB semiempirical method \citep{19BaEhGr} to obtain a more robust starting point for our following conformational search and quantum chemistry calculations. The GFN2-xTB method has proved appropriate for the fast yet reliable calculation of chemical structures. Here we excluded molecules failing to converge during the geometry optimisation process (\textit{checkpoint No. 2}).

\vspace{1em}
\noindent\textbf{Step 3: Conformational Search} \newline
Flexible molecules populate conformational states that include local and global energy minima, all having different impacts on the molecule's physico-chemical properties \citep{20PrBoGr,21GrBoHa, 22BuHaGr}, including their vibrational and rotational spectroscopy. The local and global minima correspond to different spatial geometries in which the bond order between the constituent atoms is constant. The geometries are interconvertible by rotations around single bonds. Here, we constrained our dataset to consider ground-state geometries (i.e., global energy minima) only because they correspond to the most populated molecular state at room-like temperatures. The neglected conformers are likely to be more energetically relevant at higher temperatures and for larger, less rigid molecules. 

We used the CREST \citep{20PrBoGr} and CENSO \citep{21GrBoHa} algorithms to explore the conformational space of each molecule to find the true ground-state molecular geometry. Specifically, we first input the optimised GFN2-xTB geometries from the previous step into the CREST algorithm to obtain a set of conformational geometries (conformers) and corresponding energies. The set of conformers is limited by a set of three pre-set thresholds: (1) the root-mean-square deviation of the conformer geometries (\(<\)\,0.125\,\AA), (2) the difference in the rotational constants (\(<\)\,15.0\,MHz), and (3) the energy difference between conformers (0.1\,kcal mol\(^{-1}\)). The calculations were run in gas-phase at 298.15\,K \citep{20PrBoGr}.

The resultant ensemble of conformational geometries and energies for each molecule in our working set were then input into the CENSO algorithm and refined in a four-step process \citep{21GrBoHa}. The first step improves upon the CREST electronic energy calculation with a calculation at the B97-D3/def2-SV(P) model chemistry \citep{10GrAnEh,05WeAh}, discarding high-lying energy conformers (local minima). Second, a higher level of theory (r2scan-3c \citep{21GrHaEh}) is employed to further refine the electronic energy and perform a second re-ranking and filtering of the remaining conformers while taking the computed free energies into account. Third, the remaining conformer geometries are optimised using the r2scan-3c level of theory in an 8-step optimisation process. We set \texttt{crestcheck\,\(=\)\,`on'} to remove conformer duplicates that arise after the higher-level geometry optimisation. Finally, the energies are further refined, sorted, and filtered using the PW6B95/def2-TZVPD model chemistry \citep{05ZhTr,05WeAh}. 

All CREST and CENSO calculations were performed using the ORCA 5.0.3 \citep{20NeWeBe} quantum chemistry package. We limited the walltime for the calculations to 4 hours with 1 core and 4\,Gb of memory. In \textit{checkpoint No. 3} we looked for molecules failing at either the CREST or CENSO steps and removed them from our approach.

\begin{sidewaystable}
    \centering
    \caption{Sample of the master file collating the calculated data for all molecules in the \harmdsname{} dataset of potential biosignatures, with descriptions of the column headings. The full file can be found in the supplemental material for this paper.}
    \label{tab:data_sample}
    \scalebox{0.85}{
    \begin{tabular}{p{0.05\linewidth}  p{0.13\linewidth}  p{0.07\linewidth}  p{0.01\linewidth} p{0.07\linewidth} p{0.07\linewidth} p{0.06\linewidth} p{0.04\linewidth} p{0.04\linewidth} p{0.04\linewidth} p{0.04\linewidth} p{0.04\linewidth} p{0.04\linewidth} p{0.04\linewidth}}
    \toprule
    \multirow{2}{*}{Formula} & \multirow{2}{*}{Name} & \multirow{2}{*}{SMILES} & \multirow{2}{*}{At.} & \multirow{2}{*}{Harmo} & \multirow{2}{*}{Pred. Fund.} & \multirow{2}{*}{Ints}    & \multicolumn{3}{c}{$\mu$ Components}    & \multirow{2}{*}{$\mu_{tot}$} & \multicolumn{3}{c}{Rot. Constants} \\
      \cmidrule(r){8-10}       \cmidrule(r){12-14}  
     &       & &      && &  &  $\mu_{a}$     & $\mu_{b}$     & $\mu_{c}$      &      &  A    & B     & C       \\
    \midrule
    \ce{CO}  & Carbon monoxide       & [C-]\#[O+]      & 2    & [2209.851]     & [2129.9]& [81.2]   & 0.1    & 0.00  & 0.00   & 0.1  & 0.00  & 1.93  & 1.93    \\
    & & & & & & & & & & & & & \\
    \ce{H2O} & Water & O       & 3    & [1632.1, 3820.4, ...]  & [1598.0, 3682.2, ...]   & [76.3, 3.7, ...] & 0.00   & -1.84 & 0.00   & 1.84 & 27.3  & 14.4  & 9.42    \\
    & & & & & & & & & & & & & \\
    \ce{C2H2}& Ethyne& C\#C    & 4    & [639.7, 639.7, ...]    & [636.8, 636.8, ...]     & [0.0, 0.0, ...]  & 0.00   & 0.00  & 0.00   & 0.00 & 0.00  & 1.19  & 1.19    \\
    & & & & & & & & & & & & & \\
    \ce{C2N2O}       & Cyanic isocyanate     & N\#CN=C=O       & 5    & [126.3, 468.9, ...]    & [125.7, 466.7, ...]     & [4.21, 16.1, ...]& -2.59  & -0.44 & 0.00   & 2.63 & 3.29  & 0.00  & 0.09    \\
    & & & & & & & & & & & & & \\
    \ce{H4N2}& Hydrazine     & NN      & 6    & [432.8, 813.3, ...]    & [430.8, 809.5, ...]     & [35.4, 59.1, ...]& 0.00   & 0.00  & -1.92  & 1.92 & 4.82  & 0.81  & 0.81    \\ 
    \bottomrule
    \end{tabular}}

    \scalebox{0.85}{
    \begin{tabular}{ccl}
    \\
    Column & Notation &  \\
    \midrule
    1  & Formula     & Molecular formula \\
    2  & Name& IUPAC molecar name \\
    3  & SMILES      & SMILES identifier \\
    4  & At. & Total number of atoms including both hydrogen and non-hydrogen atoms \\
    5  & Harmo       & List of raw harmonic frequencies (in \cm{}) \\
    6  & Pred. Fund. & List of predicted fundamental frequencies (in \cm{}) after using the scaling factors reported in \cite{23ZaMc} \\
    7  & Ints& List of calculated transition intensities (in km/mol) \\
    8  & $\mu_{a}$   & Dipole moment component along the $a$-axis (in $D$) \\
    9  & $\mu_{b}$   & Dipole moment component along the $b$-axis (in $D$) \\
    10 & $\mu_{c}$   & Dipole moment component along the $c$-axis (in $D$) \\
    11 & $\mu_{tot}$ & Total molecular dipole moment (in $D$) \\
    12 & A   & $A$ rotational constant (in \cm{}) \\
    13 & B   & $B$ rotational constant (in \cm{}) \\
    14 & C   & $C$ rotational constant (in \cm{}) \\
    \bottomrule
    \end{tabular}}
    
\end{sidewaystable}

\vspace{1em}
\noindent\textbf{Step 4: Harmonic Frequency Calculations} \newline
Once obtained the ground-state geometries for all successful potential biosignatures, we ran harmonic vibrational frequency calculations using the B97-1/def2-TZVPD model chemistry as recommended in \citet{23ZaMc}, due to its superior performance and affordable computational timings. Note that the basis set (def2-TZVPD) has been augmented with diffuse functions on both hydrogen and non-hydrogen atoms to achieve superior vibrational intensity predictions through enhanced dipole moment calculations \citep{20ZaMc}.

During the harmonic vibrational frequency calculations, all initial molecular geometries were optimised using an ultrafine integration grid (99 radial shells and 590 angular points per shell) and a tight convergence criterion with a maximum force and maximum displacements smaller than 2.0x10$^{-6}$\,Hartree/Bohr and 6.0x10$^{-6}$\,\AA, respectively. All calculations were carried out using the Gaussian quantum chemistry package \citep{g09d01}. A template input file of our calculations can be found in the supplemental material for this paper.

Common errors in this step (\textit{checkpoint No. 4}) included convergence issues during the geometry optimisation. Note that changing the input geometry in this step was forbidden to avoid using geometries different to the ground-state one. Only molecules with successful ground-state conformer vibrational frequency calculations are processed and analysed in this paper.

Finally, following good-practice procedures in harmonic frequency calculations \citep{21ZaMc,22ZaMc_VIBFREQ,23ZaMc}, we scaled all raw harmonic frequencies to calculate the corresponding predicted fundamental frequencies. Particularly, we used three different scaling factors optimised for the low- (0.995 for frequencies $<$\,1,000\,\cm{}), mid- (0.9792 for frequencies between 1,000--2,000\,\cm{}), and high-frequency (0.9638 for frequencies $\geq$\,2,000\,\cm{}) spectral regions to achieve superior wavenumber accuracy. \Cref{tab:mc_calcs} in Section \ref{sec:QC} presents the statistical metrics describing the anticipated accuracy of our predicted fundamental frequencies.

\subsection{Success Rate}

Out of the \totmol{} potential biosignatures initially considered, at the end of our high-throughput approach we obtained approximate vibrational spectral data for \calcmol{}  molecules (94\,\%) without requiring further manual handling (i.e., all \calcmol{} molecules successfully passed the four checkpoints described in \Cref{fig:HT_Approach}). In \Cref{table:methodreliability} we present a breakdown of the success rate for the different steps in our high-throughput approach.

\begin{table}
    \centering
    \caption{Total number of molecules with failed calculations at each step in our high-throughput workflow. Number of vibrational frequency calculation failures pertains only to the groundstate conformer species. Percentages are calculated from the total number of 2914 molecules chosen from the \citet{16SeBaPe} biosignature list.}
    \begin{tabular}{clcc}
    \toprule
Step & \multicolumn{1}{l}{Description}  & \# Successful & \# Failed \\
    \midrule
1&SMILES Geometry   & 2910 & 4     \\
2&GFN2-xTB Geom. Opt.  & 2910 & 0     \\
3&Conformer Search & 2872 & 38     \\
4&Harmonic Calculation & 2743 &  129    \\
    \midrule
& Total & 2743 (94\,\%) & 171 (6\,\%)     \\
    \bottomrule
    \end{tabular}
    \label{table:methodreliability}
\end{table}

Most failed calculations in our approach (129/171) occur at Step 4 due to geometry convergence errors during the harmonic frequency calculations. Interestingly, no imaginary frequencies were found in any of the calculations, though previously observed in \citet{23ZaMc}. 38 calculations failed at Step 3, due to either topology changes during the CREST calculation (30) or termination errors during the CENSO refinement (8). We only had 4 molecules failing to generate initial geometries from their SMILES identifiers (Step 1).

Given the nature of our automated high-throughput approach, we did not pursue alternative routes to fix the calculations broken throughout the different steps (see \Cref{table:methodreliability}). However, it is worth-noting that manual interventions could be applied to fix these calculations. We are currently assessing these issues and will update the \harmdsname{} dataset with these new data in the near future.

Specifically, for calculations failing during Steps 1, 2, and 3 (see \Cref{fig:HT_Approach}), all involving geometry-related issues, we can obtain alternative initial geometries via molecular editors (e.g., Avogadro \citep{12HaCuLo}) or cheap quantum chemistry geometry optimisations (e.g., using the HF/6-31G* model chemistry). Other errors related to topology changes during the CREST/CENSO step (Step 3 in \Cref{fig:HT_Approach}) could also be fixed using different keywords (e.g., \texttt{gfnff} for fixed input geometry and topology) in the input file that address this problem.

Fixing calculations that fail during the harmonic frequency calculation step (Step 4 in \Cref{fig:HT_Approach}) adds an extra layer of complexity because changes to the initial molecular geometry (which is the standard procedure when dealing with geometry convergence issues) could imply using a geometry different to that of the ground-state structure. Potential ways for overcoming this issue include loosening the geometry optimisation convergence thresholds in the calculations, or using a different level of theory and/or basis set (perhaps a double instead of triple-zeta augmented basis set). However, we currently have no evidence of success when implementing these solutions to the failed harmonic frequency calculations.

We make the assumption that molecules that have successfully passed Step 4 have ground-state geometries.

\subsection{Description of Format}

The \harmdsname{} calculated vibrational spectral data can be found in the supplemental material for this paper. \Cref{tab:data_sample} presents a sample of the master file collating all calculated frequency data along with a brief description for each column in the file.

\section{Vibrational Spectroscopy Data}
\label{sec:vib_spec}

\begin{figure*}
    \centering
    \includegraphics[width = 1\textwidth]{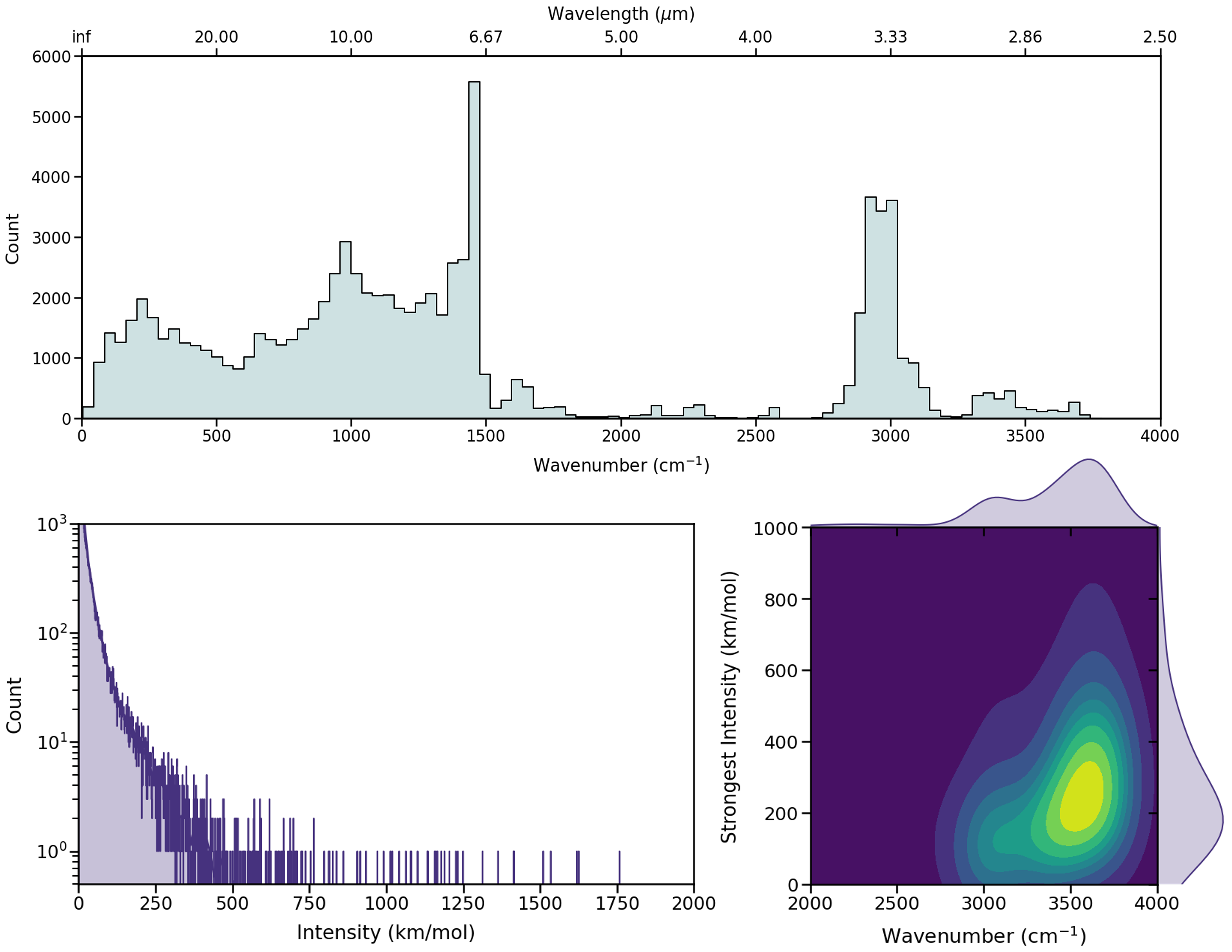}
    \caption{Top: Predicted fundamental frequency distribution for all molecules in the \harmdsname{} dataset of potential biosignatures. Please note that this plot is not a spectrum but a histogram of the all calculated frequencies. Bottom-left: Distribution of calculated vibrational transition intensities.  Bottom-right: Comparison between the strongest vibrational transition intensity per molecule with its corresponding predicted fundamental frequency. Note that only intensities with up to 500\,km/mol and frequencies above 2000\,\cm{} are consider for the sake of readability.}
    \label{fig:VibFreq_Data}
\end{figure*}

This section provides an overall description of the approximate vibrational spectral data produced here and available within the \harmdsname{} dataset. Specifically, we delve into the distribution of the predicted fundamental vibrational frequencies and intensities for all molecules, and make some individual comparisons against experimental values.

\subsection{Frequency and Intensity Distribution}

\Cref{fig:VibFreq_Data} presents an overview of our predicted fundamental frequencies and transition intensity  for all molecules in the \harmdsname{} dataset. The top figure shows a histogram of the predicted fundamental frequencies, the bottom left shows the distribution of predicted transition intensities, while the bottom right plot correlates the strongest intensity per molecule with its corresponding frequency. There is significant information in these figures that is worth exploring in depth. 

The top plot in \Cref{fig:VibFreq_Data} presents a histogram of the spectral distribution of all predicted fundamental frequencies across the \calcmol{} molecules in the \harmdsname{} dataset. Please note that this plot is not a spectrum. The large number of frequencies near 3000\,\cm{} corresponds to the C--H stretch vibrational mode, which correlates with the dominance of hydrocarbons in our dataset (see \Cref{fig:fgroups}). Other common vibrational modes that highlight in the plot include the C=O, N--H, and O--H stretches, lying at approximately 1700, 3300, and 3600\,\cm{}, respectively. The high count of frequencies near 1500\,\cm{} can be attributed to the C--H bending modes. 

Our calculated quantum chemistry data also provide a good description of the influence of the chemical environment on functional group vibrations (a key limitation for the RASCALL approach). The impact of chemical environment on functional group frequencies is evident in the top plot in \Cref{fig:VibFreq_Data} where the central frequencies for common functional group vibrations (e.g., C--H stretch at 3000\,\cm{}) span a wide range of frequencies, rather than of a single frequency value. Our data predict a significant difference of 70\,\cm{} (experimentally 80\,\cm{} \citep{57GrLo,01KiZe}) in the symmetric C--H stretch frequency of methylamine (\ce{CH3NH2}) and methylphosphine (\ce{CH3PH2}), despite their structural similarities due only to the replacement of the nitrogen by a phosphorus atom.

Looking at the low-frequency fingerprint region (below 1000\,\cm{}), it is clear that our computational quantum chemistry approach addresses another key limitation of RASCALL by enabling non-local modes to be naturally predicted. Having data in this region could play a critical role in (1) disentangling the identity of molecules with similar functional groups and high-frequency spectra (vibrational modes in the fingerprint region mostly correspond to coupled vibrations involving large parts of the molecule, which makes them unique to individual molecules), and (2) allowing detection of molecules with weak high-frequency bands. For instance, the detection of \ce{C2N2} on Titan's atmosphere \citep{81KuAiHa} was possible thanks to observations of the 220\,\cm{} band as the higher frequency vibrations for this molecule have virtually null intensities. These low-frequency spectral windows can be utilised by JWST \citep{patapis2022direct} with the mid-infrared instrument (MIRI) which can measure out to 28.3\,\um (<\,400\,\cm{}). Though further exploration is needed, if it is determined that this spectral region is particularly useful for understanding exoplanet atmosphere compositions, this evidence can be used to motivate the design of future instruments.

It is worthwhile noting the low number of vibrational modes with predicted fundamental frequencies in the region between 1500--2800\,\cm{}. This region of lower spectral congestion is likely more useful for definitive molecular detection than the crowded 3000--3500\,\cm{} and 1000--1500\,\cm{} spectral windows. 

The bottom-left plot in \Cref{fig:VibFreq_Data} presents the logarithmic-scale distribution of all transition intensities in \harmdsname{}. More than 95\,\% of the computed values lie below 500\,km/mol , with numbers drastically decreasing at much larger intensity values. 
Note that weak transitions (below 500\,km/mol) can be observed for molecules with  high abundances, e.g., \ce{OH} \citep{21NuKaGi}, \ce{H2O} \citep{07TiMaLi,13DeWiMc,14KrBeDe,14LoJoBe,15KrLiBe,23AlWaAl}, \ce{NH3} \citep{21GiBrGa}, and \ce{CH4} \citep{08SwVaTi,11BaMaKo,14StBeMa,15BaKoMa}, all with confirmed atmospheric detections on (exo)planetary atmospheres. 

The bottom right-side plot in \Cref{fig:VibFreq_Data} correlates the strongest intensity per molecule with its corresponding frequency. The frequency window is restricted because almost all molecules have their strong spectral features in the region between 3000 and 4000\,\cm{}, which is the frequency range dominated by C--, N--, and O--bearing stretching modes. This spectral region is thus the most useful in terms of detecting signals for new molecules; however, definitive identification of a single molecule will be difficult given the large number of molecular absorbers in this region. 

\begin{figure*}
    \centering
    \includegraphics[width = 1\textwidth]{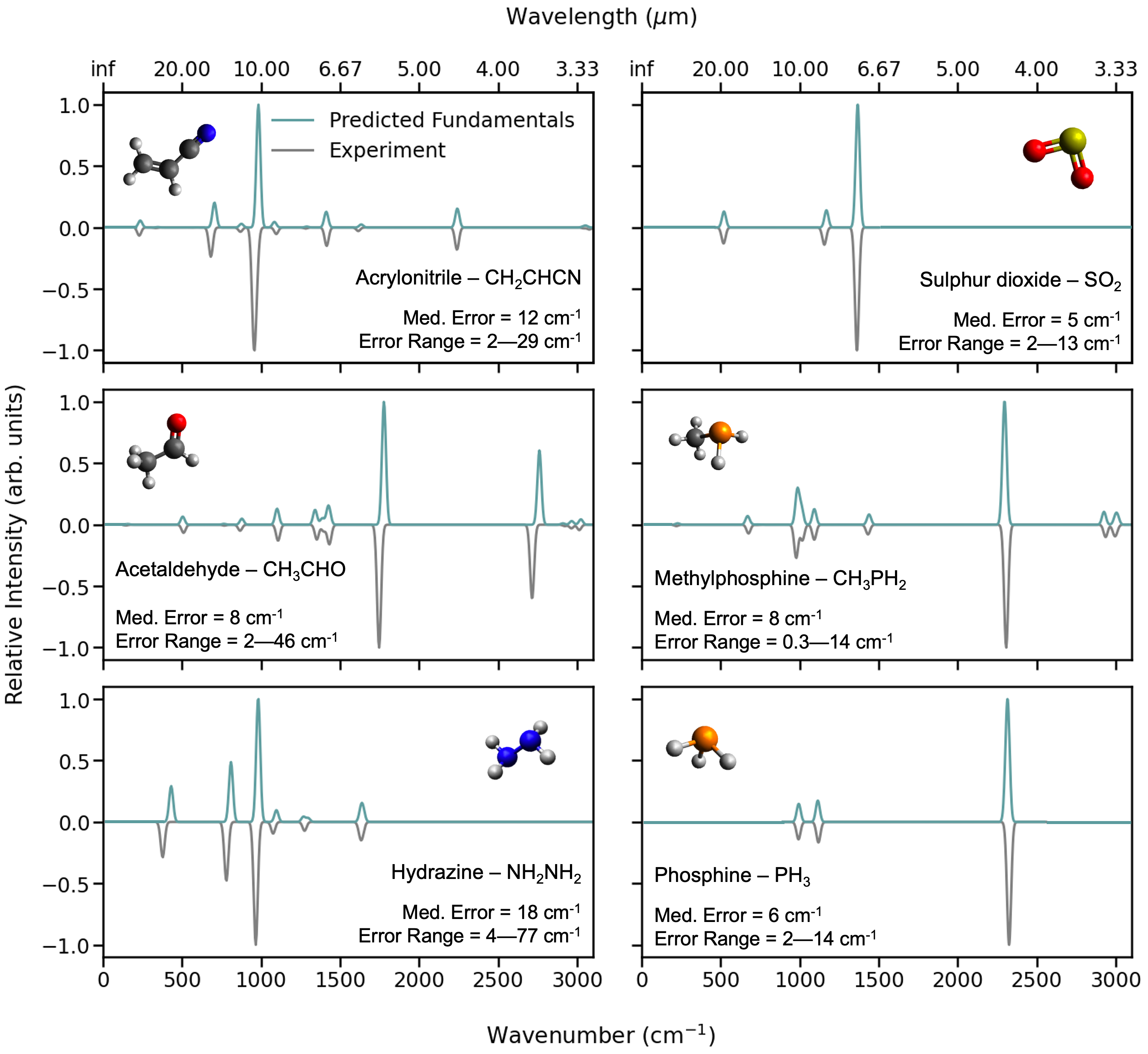}
    \caption{Comparison between our predicted fundamental frequency data and the experimental data available for acrylonitrile (\ce{CH2CHCN}) \citep{99KhNoPa,15KiMaPi}, sulphur dioxide (\ce{SO2}) \citep{87GuNaUl,05MuBr}, acetaldehyde (\ce{CH3CHO}) \citep{71HoGu,80Ho,94KlHe,95AnTrPa,99HeHeGe}, methylphosphine (\ce{CH3PH2}) \citep{01KiZe}, hydrazine (\ce{NH2NH2}) \citep{52GiLi,60YaIcSh,75DuGrMa}, and phosphine (\ce{PH3}) \citep{92TaLaLe,00FuLo}. An artificial Gaussian profile has been added to the central frequency peaks from both the experimental and quantum chemistry data for illustrative purposes only; please note the true spectra would have a considerably different rovibrational profile.}
    \label{fig:QCExp_Comp}
\end{figure*}

\subsection{Individual Molecule Comparisons}

\Cref{fig:QCExp_Comp} compares the experimental and predicted fundamental frequency data (calculated from our approach) for six different molecules in the \harmdsname{} dataset. Overall, the agreement is strong, with median errors between 5 to 18\,\cm{}, in line with the expectations from the \cite{23ZaMc} benchmarking. Hydrazine and acetaldehyde have the  most significant outliers with maximum deviations of 77 and 46\,\cm{} respectively. The presence of some quite significant outliers is not unexpected based on benchmarking results, and is a known limitation of our approach to predicting fundamental frequencies by scaling harmonic frequencies generated through DFT methods. 

\subsection{Comments on Data Utilisation in Astrochemistry} 

In analysis of JWST-resolution data, astronomers must be aware that many molecules have similar spectra, leading to potential incorrect assignments if only relying on available line lists and small spectral windows. The \harmdsname{} allows identification of other potential assignments for unknown spectral features, useful for unexpected exoplanetary chemical compositions. However, there are valid reasons for selecting the molecules for which line lists are available, including previous astronomical detections in cool stars and solar system bodies and sophisticated chemical modelling of exoplanetary atmospheres. 

The transition intensity data in \harmdsname{} can identify molecules detectable at low abundance, but other factors like clouds, hazes, and stellar contamination also need to be considered \citep{07ShBu,22NiWGo,20BaHe,18RaApGi}.

\section{Applications in Astrochemistry: WASP-39\lowercase{b} \ce{SO2} Atmospheric Detection}
\label{sec:astrochem}

The prime application of our quantum chemistry big data is aiding in the identification of molecules in spectra recorded from different astronomical sources. Here we provide a key example showcasing how our data provide crucial insights into paving the way to unveiling new potential molecular absorbers to unknown spectral features.

Recent single-transit observations of WASP-39b included a spectral band initially unidentified at 4.1 \um{}. 
Acknowledging that \ce{SO2} is now confidently identified as the origin of the 4.1 \um{} absorption \citep{23AlWaAl}, here we present how our quantum chemistry big data could have highlighted \ce{SO2} as a potential molecular candidate for the 4.1\,\um{}  ``unknown" spectral signal, should the detection not yet be confirmed. Specifically, we use this example to highlight how our data can play a pivotal role in screening potential molecular candidates to unknown spectral lines in the future.

\subsection{Data Set Constrains}

\begin{table*}
    \centering
    \caption{Molecules from the \harmdsname{} dataset with fundamental frequency absorption (in \cm{} and \um{} ) between 2320 and 2620\,\cm{}, along with their SMILES identifiers and calculated transition intensities (in km/mol). When available, the corresponding experimental fundamental frequency is provided with its corresponding reference to the original publication.}
    \label{tab:UB_Fundamentals}
    \scalebox{0.8}{
    \begin{tabular}{llcccccc}
    \toprule
\multirow{2}{*}{Formula}      & \multirow{2}{*}{SMILES Identifier}   & \multicolumn{2}{c}{Predicted Fundamental Frequency}  & \multirow{2}{*}{Intensity} & \multicolumn{2}{c}{Experimental Frequency} & \multirow{2}{*}{Reference}  \\
\cmidrule(r){3-4} \cmidrule(r){6-7}
      &      & \cm{}& \um{}&    & \cm{} & \um{}      &     \\
    \midrule
\ce{C2HNS2}   & SC(C\#N)=S   & 2549 & 3.92 & 5.97       & -     & -  & -   \\
\ce{HSCN}     & SC\#N& 2544 & 3.93 & 4.86       & 2580.9& 3.87       & \citep{01WiMi}       \\
\multirow{2}{*}{\ce{H2S}}     & \multirow{2}{*}{S}   & 2578 & 3.88 & 0.04       & 2614.4& 3.82       & \citep{98BrCrCr}     \\
      &      & 2591 & 3.86 & 0.05       & 2628.5& 3.8& \citep{98BrCrCr}     \\
\multirow{2}{*}{\ce{H3O2P}}   & \multirow{2}{*}{O=[PH2]O}    & 2333 & 4.29 & 91.01      & -     & -  & -   \\
      &      & 2366 & 4.23 & 67.94      & -     & -  & -   \\
\ce{H3OP}     & [H][P]([H])([H])=O   & 2327 & 4.30 & 73.23      & -     & -  & -   \\
\ce{HNO2S}    & [O-][N+](=O)S& 2569 & 3.89 & 4.01       & -     & -  & -   \\
    \bottomrule
    \end{tabular}}
\end{table*}

The 4.1\,\um{} (2439\,\cm{}) spectral band extended from 3.9--4.2\,\um{}  (2380--2564\,\cm{}). Using the the 95\,\% percentile error range as an upper bound on the expected accuracy of our predicted fundamental frequencies, we constrained our search to molecules with predicted frequencies between 2320 and 2620\,\cm{} (3.8--4.3\,\um{}).

Importantly, the infrared spectrum of WASP-39b has no strong spectral features near 3.03 (3300\,\cm{}) and 3.33\,\um{} (3000\,\cm{}), corresponding to the central frequency for the N--H and C--H stretching vibrational modes, respectively. The absence of these stretching modes means we can infer there are no N--H and C--H bonds in the molecule responsible for the 4.1\,\um{} (2439\,\cm{}) feature. 
Therefore, using the RDKit library in python, we excluded all molecules containing N--H and C--H groups from our frequency search.

Finally, as WASP-39b is a hot gas-giant exoplanet (T$_{ex}$ = 1120\,K), we constrained our frequency search to only molecules with less than or equal to 6 total atoms including H; larger molecules are unlikely to be stable at this temperature. 

\subsection{Fundamental Frequency Search}

The sparsity of our selected spectral window and tight molecular constraints meant that only six molecules from \harmdsname{} have predicted fundamental frequencies in this region.  These molecules and their corresponding frequencies are detailed in \Cref{tab:UB_Fundamentals}. Though we can highlight some interesting candidates, most molecules in the table are  unusual and don't have  spectroscopic data available in the literature.
 
Hydrogen sulfide (\ce{H2S}) and thiocyanic acid (\ce{HSCN}) stand out as the two most interesting molecular candidates for the 4.1\,\um{} (2439\,\cm{})  spectral signal as they have both been observed in the interstellar medium (ISM) \citep{72ThKuPe,09HaZiBr} and they are relatively simple molecules more likely to be present in the atmospheric conditions of WASP-39b. However, the high-resolution infrared spectral data available for both molecules (experimental frequencies in the table) shows that the actual fundamental frequencies were much higher than the 4.1\,\um{} (2439\,\cm{}) line in the spectrum of WASP-39b, leading us to exclude these molecules as potential candidates for this signal. 

The absence of any feasible candidates within our \harmdsname{} dataset for the 4.1\,\um{} (2439\,\cm{}) spectral feature strongly suggest that the signal does not correspond to a fundamental vibrational frequency absorption. This motivated our investigation of the overtones and combination bands in search of alternative potential molecular candidates.

\subsection{Beyond fundamentals}

One of the significant limitations of our \harmdsname{} dataset is that it only includes predicted fundamental frequencies. It is plausible to use these to predict overtone and combination bands, but from a practical point of view it is easier and more reliable to instead perform more expensive VPT2 calculations only for molecules within our constraints (six or fewer atoms and no C--H and N--H bonds).

Thus, for this small subset of molecules (77 molecules in total), and following the same high-throughput procedure described in Section \ref{sec:approach}, we used the molecular geometries from the \harmdsname{} dataset to perform VPT2 calculations using the B97-1/def2-TZVPD model chemistry.  A template of the input file for running the VPT2 anharmonic calculations is provided in the supplemental material for this paper. 

We note that, without benchmarking this methodology against experiment, we don't have a strong understanding of the errors. While the performance of our model chemistry for harmonic frequency calculations could be a good indication of upper bound performance, previous studies have found that some DFT methods struggle with the higher-order derivatives involved in VPT2 calculations leading, in some cases, to startling poor errors of hundreds to thousands of wavenumbers \citep{20BaCeFu,21ZaSyRo}. Therefore, an anharmonic benchmark study outlining errors and anticipated performance is a high priority for further study. 

We identified 18 candidate molecular species with at least one overtone and/or combination band within the 2320--2620\,\cm{} frequency window (3.8--4.3\,\um{}). 
\Cref{tab:Overtones_ComBands} presents the most likely of these candidate molecules, with their corresponding calculated frequencies and transition intensities; the full version of this table is presented in the supplemental material. 

\begin{table}
    \centering
    \caption{Selection of candidate molecules for the 4.1 \um{} spectral feature based on the presence of overtone and/or combination bands with frequencies between 2320 and 2620\,\cm{}. The molecule formula, SMILES identifier, frequency of each mode (in \cm{} and \um{} ) and calculated transition intensities (in km/mol) are tabulated.}
    \label{tab:Overtones_ComBands}
    \scalebox{1}{
    \begin{tabular}{llccHc}
    \toprule
\multirow{2}{*}{Formula}    & \multirow{2}{*}{SMILES }& \multicolumn{2}{c}{Calc. Frequency} & & \multirow{2}{*}{Intensity} \\
\cmidrule(r){3-4}
    && \cm{} & \um{} &       &  \\
    \midrule
\multirow{3}{*}{\ce{C2N2O}} & \multirow{3}{*}{N\#CN=C=O}     & 2390  & 4.18  & Combination Band      & 1.6\\
    && 2431  & 4.11  & Combination Band      & 2.5\\
    && 2482  & 4.03  & Combination Band      & 26.5       \\ [2mm]
\multirow{6}{*}{\ce{C2N2S}} & \multirow{6}{*}{N\#CN=C=S}     & 2379  & 4.20  & Combination Band      & 27.3       \\
    && 2448  & 4.09  & Overtone      & 63.8       \\
    && 2533  & 3.95  & Combination Band      & 0.2\\
    && 2551  & 3.92  & Combination Band      & 6.5\\
    && 2587  & 3.87  & Combination Band      & 0.1\\
    && 2588  & 3.86  & Combination Band      & 0.4\\ [2mm]
\multirow{2}{*}{\ce{H2O2}}  & \multirow{2}{*}{OO}    & 2333  & 4.29  & Combination Band      & 0.07       \\
    && 2545  & 3.93  & Overtone      & 0.5\\ [2mm]
\multirow{5}{*}{\ce{HNO3}}  & \multirow{5}{*}{O[N+]([O-])=O} & 2361  & 4.24  & Combination Band      & 0.1\\
    && 2484  & 4.02  & Combination Band      & 0.001      \\
    && 2553  & 3.92  & Overtone      & 2.3\\
    && 2601  & 3.84  & Combination Band      & 3.8\\
    && 2609  & 3.83  & Combination Band      & 0.7\\ [2mm]
\ce{HOP}    & P=O    & 2473  & 4.04  & Overtone      & 0.7\\ [2mm]
\multirow{3}{*}{\ce{O3}}    & \multirow{3}{*}{O=[O+][O-]}    & 2425  & 4.12  & Overtone      & 0.1\\
    && 2458  & 4.07  & Combination Band      & 8.0\\
    && 2513  & 3.98  & Overtone      & 0.010      \\ [2mm]
\multirow{2}{*}{\ce{SO2}}   & \multirow{2}{*}{O=S=O} & 2367  & 4.23  & Overtone      & 0.29       \\
    && 2564  & 3.90  & Combination Band      & 3.15       \\

    \bottomrule
    \end{tabular}}
\end{table}

\begin{figure}
    \centering
    \includegraphics[width=0.8\textwidth]{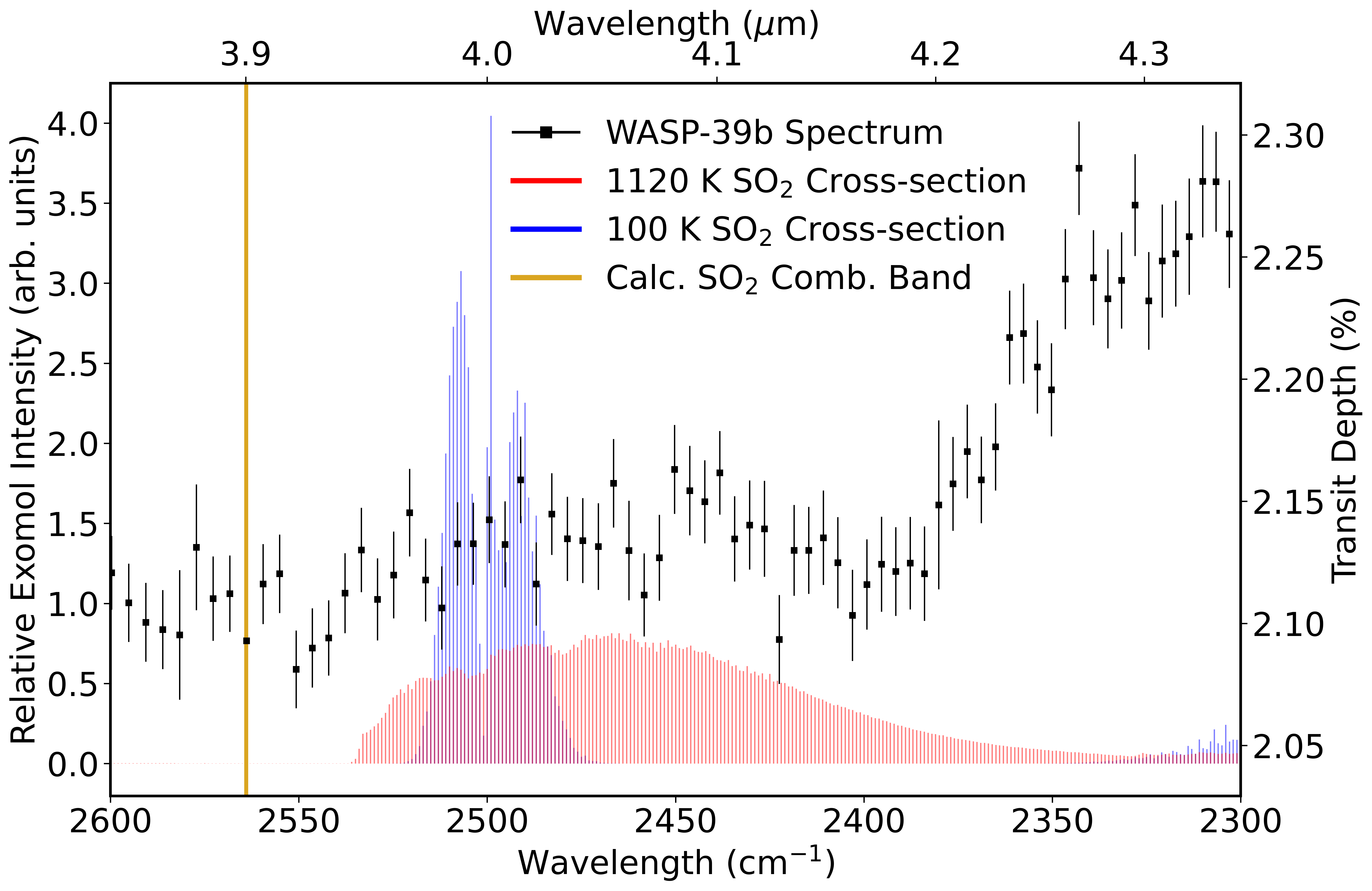}
    \caption{Detailed investigation into the \ce{SO2} combination band in the context of the WASP-39b (atmospheric temperature of 1120\,K). Comparison between the JWST observational data (black), the VPT2-calculated combination band centre (yellow) at 2564\,\cm{} and the cross-section at 100\,K and 1120\,K modelled from the ExoMol \ce{SO2} line list using the online ExoMol absorption cross-section service \citep{OnlineXsec}. }
    \label{fig:JWST_SO2}
\end{figure}

Reassuringly, our quantum chemistry big data confirms \ce{SO2} as one of the candidate molecules. Its strongest absorption in the spectral region is predicted as the combination band at 2564\,\cm{} (3.90\,\um{}), with another weaker overtone absorption at 2367\,\cm{} (4.23\,\um{}) that is within the strong \ce{CO2} band present in the WASP-39b atmospheric spectrum. However, there is a substantial difference ($>$\,100\,\cm{}) between the predicted band centre (2564\,\cm{}, 3.9\,\um{}) and the observed band peak (2439\,\cm{}, 4.1\,\um{}). We investigate this further in \Cref{fig:JWST_SO2}, which compares the JWST atmospheric spectrum of WASP-39b against the position of our predicted central band for \ce{SO2} (yellow vertical line) and the ExoMol \ce{SO2} cross-sections \citep{SO2_ExoMol} at 100\,K and 1120\,K (the temperature of WASP-39b). This comparison makes it clear that there are two main causes behind the large deviation between our predicted frequency and the frequency observed in the spectrum: 
\begin{enumerate}
    \item the computational prediction of the combination band centre frequency is in error by approximately 60\,\cm{} (this is evident due to the large discrepancy between the large central Q-band peak of \ce{SO2} at 100 K and our calculated \ce{SO2} combination band frequency); and
    \item at high temperatures, the peak of the rovibrational band profile is shifted by about 50\,\cm{} (this is shown by the position of the band peak, i.e., the maximum band intensity, of the 100\,K and 1120\,K predicted cross-section data).
\end{enumerate}

Follow-up of other candidate molecules is beyond the scope of this paper, but feasible given the small number.  

\subsection{Discussion}

By using high-throughput quantum chemistry data to investigate the 4.1\,\um{} (2439\,\cm{}) spectral line  in the WASP-39b infrared spectrum (now known to be a combination band of \ce{SO2}), we can identify the strengths and weaknesses of this approach, and prioritise areas for future improvements. 

By considering approximate fundamental, overtone and combination band frequencies for small molecules without C--H and N--H bonds, it was straightforward to produce a modest set of molecular candidates for the 4.1\,\um{} (2439\,\cm{}) signal that included \ce{SO2}. However, excluding molecules based solely on the match between the astronomically observed band peak and the experimental band centre was premature, as the rotational profile of the band can cause a significant difference between the band centre (the vibrational-only transition) and band peak (the frequency with highest intensity once rotational population and excitation are considered), particularly at higher temperatures. 

The neglect of the rotational profile and unexpectedly high error in the predicted combination band centre meant that the wide spectral window was crucial in ensuring \ce{SO2} was one of our candidate molecules. We were lucky to find only 18 potential molecular candidates in this wide spectral region. This small number can be attributed to the strong constraints on molecule type (i.e. no N-H or C-H bonds) and the sparsity of the frequency range considered; this cannot be expected for typical unidentified spectral signatures. In the future, reducing the error in the anharmonic calculations and modelling temperature-dependent rotational profiles of bands will be crucial in allowing smaller spectral windows.

Thorough benchmarking of anharmonic calculation performance will allow the selection of a model chemistry with smaller, well-characterised errors. Benchmarking anharmonic calculations of fundamental frequencies is straightforward and should follow the approach of \cite{23ZaMc}.  However, benchmarking the quality of overtone and combination band frequency predictions will be more challenging as an experimental dataset will need to be constructed or very high-level theoretical calculations performed (both are done for fundamental vibrational frequencies in \cite{22ZaMc_VIBFREQ}). Once this benchmarking is performed, we can expect that for fundamental frequencies a median error of 5\,\cm{} and/or a significant reduction in the number of outliers (quantified by the 95\,\% error range) to be achievable. It is currently unclear the accuracy to be expected for overtone or combination band modes. The observed large combination band frequency error for \ce{SO2} is either an outlier result or an indication that overtone and combination bands will have higher errors than fundamental bands.  

When considering producing temperature-dependent rotational profiles, currently, our reported data do include rotational constants that can be used manually to produce approximate rovibrational spectra for individual molecules of interest. Automating this process to produce predicted rovibrational absorption cross-sections for molecules is an active area of further development. 

Though not done extensively for this paper given the strength of the \ce{SO2} assignment, with a sufficiently small set of potential candidates for an unknown spectral signal, manual follow-up could include tracking down (or potentially producing) experimental spectra, performing higher quality computational quantum chemistry calculations, and assessing chemical models for the likelihood that these species can be produced in the specific exoplanet's environment.

It is also worth noting that even with only the approximate data of this paper, we can readily consider the candidate molecule absorption cross-section across the full infrared spectral region and quickly identify ways to confirm a tentative assignment or distinguish between potential assignments with minimal follow-up observations. In the case of \ce{SO2}, the computational quantum chemistry data alone make it abundantly clear that the best way to unambiguously confirm the assignment is  to look for the fundamental vibrational frequencies of \ce{SO2} at the symmetric (1155.7\,\cm{}, 8.65\,\um{}) and asymmetric (1362.1\,\cm{}, 7.34\,\um{}) S=O stretches \citep{87GuNaUl}, both of which will be far stronger than the combination band at 4.1\,\um{}.

\section{Final Remarks and Future Directions}
\label{sec:final_remarks}

The launch of the James Webb Space Telescope (JWST) has unlocked a new and stimulating avenue to survey exoplanetary atmospheres. Molecular spectroscopic data are crucial to maximising the gains in our understanding of the molecular composition of exoplanetary atmospheres, providing insights into the planets' physical, chemical, and even potential biological processes.  High-resolution, high-completeness spectroscopic data in the form of line lists are rapidly being produced to allow these detections, but is still only available for a relatively small number of species, limiting the scope of new molecular detections in exoplanetary atmospheres.

In this work, we have produced approximate spectral data for a large number of molecules; specifically, the \harmdsname{} dataset with predicted vibrational (infrared) frequencies and intensities for \calcmol{} small molecules identified as potential biosignatures with an estimated accuracy of 10\,\cm{} (0.01--0.10\,\um{} depending on spectral region). As demonstrated here for the 4.1\,\um{} (2439\,\cm{}) line in the infrared spectrum of WASP-39b (recently identified by \cite{23AlWaAl} as \ce{SO2} using ExoMol line list data), our approximate data can be useful to identify potential molecular candidates to unknown spectral features for further follow-up studies. 

Definitive molecular detections using our quantum chemistry data alone are impossible, yet approximate spectral data for a very large number of potential absorbers fill an important gap in existing approaches. We thus anticipate our data to be useful in investigating spectral features originating from molecules for which no collated line list data are available; for example, the 4.25\,\um{} feature on the WASP-39b spectrum whose molecular origins are currently unknown \citep{23AlWaAl}. We are currently undertaking this analysis.

Some crucial and perhaps surprising things emerged from our pilot investigation of the 4.1\,\um{} (2439\,\cm{}) line; in order of importance:
\begin{enumerate}
   \item Rotational profile can cause the band maximum (frequency of maximum intensity) to be shifted significantly from the band centre (the position of the purely vibrational transition without rotational contributions); this is particularly pronounced at high temperatures. 
    \item There is far more molecular spectral data available in the literature than is readily available to astronomers; however, this data are indexed by molecule not by spectral feature and in disperse not collated form. A modest set of candidate molecules can thus be feasibly followed up manually.  
    \item Spectral windows with no absorption are as important as the unidentified spectral feature itself in making identifications.
   \item Within some narrow JWST observational windows, for some molecules the strongest absorption will be caused by overtone and combination bands; the accuracy of computational predictions of these band centres is unknown and benchmarking is hihgly required. 
\end{enumerate} 

The production of the \harmdsname{} data relied on a newly developed high-throughput quantum chemistry approach, described and piloted in this paper, which produced data in an automatic manner for 94\,\% of the molecules considered. The CREST/CENSO algorithm has been used to more reliably generate the true ground state of each molecule. The major cause for failed calculations is the non-convergence of the molecular geometry at the vibrational calculation step.  Comparison of experimental vs theoretical fundamental frequencies in this dataset corroborated the validity of the benchmarking for our selected model chemistry approximation.  

Further extensions to this high-throughput approach are needed to maximise the usefulness to astronomers seeking to identify unknown molecular absorbers: temperature-dependent rotational profiles for every vibrational band, predictions of overtone and combination bands, and inclusion of non-ground-state conformers are the highest priority extensions. 

There are a large number of other potential applications of our bulk approximate vibration spectroscopy data including: 
\begin{enumerate}
    \item identifying ambiguities in molecule assignments, i.e., two or more molecules with similar spectral signals; 
    \item quantify spectral congestion in a given frequency window; 
    \item identifying molecules with strong absorptions that could be detectable even at low abundance;
    \item parameterise the RASCALL approach, thereby allowing fast bulk predictions of molecule spectra for very large molecules beyond the size limitation of quantum chemistry methods; and
    \item training set in machine learning algorithms. Models trained on these data could either be used to predict the infrared spectra of molecules without using quantum chemistry or to infer molecular structures from unknown infrared spectra. i.e., predict a molecule's identity given a set of infrared spectral bands.
\end{enumerate}

\begin{acknowledgement}

This research was undertaken with the assistance of resources from the National Computational Infrastructure (NCI Australia), an NCRIS enabled capability supported by the Australian Government.

The authors declare no conflicts of interest. 

\end{acknowledgement}

\begin{suppinfo}

The data underlying this study are available in the published article and its online supplementary material. 

We provide the master csv file containing the calculated spectral data for all molecules in the \harmdsname{} dataset. This files contains all predicted fundamental frequencies and transition intensities using our quantum chemistry high-throughput approach, as well as dipole moments and rotational constants for all molecules. The CHNOPS2743-HARMONIC data set can also be found in the Harvard Dataverse repository at \url{https://doi.org/10.7910/DVN/0DLSDP.}

We also provide a PDF file including additional information regarding our molecular candidates to the 4.1\,\um{} \ce{SO2} signal and samples of the input files used to run the harmonic and VPT2 anharmonic calculations.

The csv file containing the VPT2 anharmonic data for the 77 small molecules in the \harmdsname{} dataset is also provided. We warn the user, however, that we have no clear understating of the errors associated with these data as no thorough benchmarking of anharmonic approaches is currently available.

\end{suppinfo}


\bibliography{references_bigdata,references_bigdata2}

\end{document}